\documentclass[conference]{IEEEtran}
\ifCLASSINFOpdf
\else
\fi
\usepackage{times}  
\usepackage{epsfig}
\usepackage[TABBOTCAP]{subfigure}
\usepackage{tabularx}
\usepackage{graphicx} 
\usepackage{color}
\usepackage{xspace}
\usepackage{thumbpdf}
\usepackage{listings}
\usepackage{verbatim}
\usepackage{hyperref}
\usepackage{booktabs}
\usepackage{colortbl}
\usepackage{xspace}
\usepackage{epsfig}
\usepackage{amssymb}
\usepackage{amsmath}
\usepackage{amsfonts}
\usepackage{listings}
\usepackage{algorithm}
\usepackage{algorithmic}
\newcommand{\INDSTATE}[1][1]{\STATE\hspace{#1\algorithmicindent}}

\newcommand{\vcell}{{\em vcell}\xspace}
\newcommand{\vcells}{{\em vcells}\xspace}

\begin{document}
%
\title{Peer provided cell-like networks built out of thin air}

\author{\IEEEauthorblockN{Andr\'es Arcia-Moret, Zafar Gilani,\\
Arjuna Sathiaseelan, Jon Crowcroft}
\IEEEauthorblockA{Computer Laboratory\\
University of Cambridge\\
Cambridge, Cambridgeshire CB3 0FD, UK\\
Email: firstname.lastname@cl.cam.ac.uk}}


%


\maketitle

\begin{abstract}

The success of WiFi technology as an efficient and low-cost last-mile access solution has enabled massive spontaneous deployments generating storms of beacons all across the globe. Emerging location systems are using these beacons to observe mobility patterns of people through portable or wearable devices and offers promising use-cases that can help to solve critical problems in the developing world. In this paper, we design and develop a novel prototype to organise these spontaneous deployments of Access Points into what we call virtual cells (\textit{vcells}). We compute virtual cells from a list of Access Points collected from different active scans for a geographical region. We argue that virtual cells can be encoded using Bloom filters to implement the location process. Lastly, we present two illustrative use-cases to showcase the suitability and challenges of the technique.

\end{abstract}


%
\IEEEpeerreviewmaketitle

\section{Introduction}
\label{sec:introduction}

Due to recent advances in technology that allow mobile devices to find geolocation fairly accurately, there has been tremendous growth of interest in developing and exploiting location-based services.  These services enable users to share their activities and happenings in real-time, since they are integrated with social networking applications. Popular examples include Facebook Places, Foursquare and LevelUp.

In this paper, we present an initial work that proposes to use the simple active scanning process in Wi-Fi to organise spontaneous deployments of Access Points (APs) in urban areas into what we call virtual cells (\vcells). A \vcell corresponds to a collection of lists of APs produced by progressive scanning on the move. Moreover, each \vcell is computed per geographical region, typically small in size or area. The size of the \vcell varies and depends on the speed of collection and the particular device performing the scanning. There are several reasons responsible for this variation such as the antenna, the radio, the transmission conditions or the full scanning time (i.e., total time spent on actively receiving beacons \cite{arciamoret1}). We discuss a mechanism that uses Bloom filters (BF) \cite{bloom} to represent \vcells. This mechanism is intended to locate conveniently mobile devices given a Bloom filter containing very few or approximate fingerprints\footnote{A \textit{fingerprint} is a list of access points collected after a single scan.}. Our intention is to rely on a preliminary process of collection of the ground-truth for geolocation by using a GPS. Then we build a compact and fast database indexed by BFs that efficiently maps \vcells into locations.

Various commercial enterprises offer many widely accessible geolocation services. For example, Google's Android API uses the location from a GPS or the network service provider to determine device's approximate location. Similarly, iOS API uses geotagged locations of nearby cellular towers or Wi-Fi hotspots, along with its crowdsourced database of Wi-Fi hotspots, to determine device's approximate location. Commercial Web services such as Google Maps Geolocation API use MAC addresses and many other metrics (e.g., IP, age, channel, signal strength, etc.) to provide a location. Our proposed technique uses fingerprints of well-known locations of Wi-Fi hotspots to approximate a device location to a reference point or a landmark. Different from mentioned services, our proposal is intended to be open source, self-contained and crowd-sourced service.

The novelty of our technique consists of (1) grouping APs found in spontaneous deployments into cell-like network topology to assist a location service, (2) whereas other location prediction techniques use passive scanning, uncontrolled active scanning, or full connection information from Layer 3~\cite{piotr,kawaguchi,rekimoto,muthukrishnan,song}, \vcells are built from controlled Layer 2 active scanning considering the dynamics described in~\cite{arciamoret1}, (3) \vcells can be duly encoded into BFs for facilitating the location retrieval, (4) while both GPS and other services mentioned above can only work outdoors, \vcells can also offer a location service for indoors.

Some of the potential use-cases for micro-mobility through \vcells are listed below.

\textbf{Need for localisation amidst disasters.} Internet connectivity during extraordinary circumstances cannot be trusted. However, we may have islands of Wi-Fi APs that can be utilised by a stand-alone application to help people locate guidance maps, safe areas, rendezvous points, etc.

\textbf{Human tagged locations.} Naming and description of the addresses are not formalised and may be subject to whims of domineering authorities in developing regions. They may also be under-reported on common services such as Google Maps, either because of lack of a business case or feasibility to deploy Google street view cars. It is also normal to have nameless streets, short-cuts, by-paths, thus making \vcells a mean to have an ad-hoc common agreement based on a mutual consensus of the crowds for approximate location.

\textbf{Mitigate the lack of cellular coverage.} With the proliferation of IoT devices and community networks that provide WiFi access, it becomes convenient to deploy APs that may just act as location anchors. They are low-cost, energy-efficient (compared to a GPS solution), and can even be deployed with a small sustainable source of energy (e.g. solar panels).

The rest of the paper is divided into following sections. Section~\ref{sec:design} describes three main components of the prototype system. We evaluate the \vcell algorithms in two illustrative journeys and propose the applicability of Bloom filters for compact representation in Section~\ref{sec:evaluation}. Section~\ref{sec:related} describes related work in this field. Finally, we conclude the paper in Section~\ref{sec:conclusion}. 



\section{System Design}
\label{sec:design}

%
The system prototype considers three main components: the mobile application, the \vcell forming algorithms and the Bloom filter based location. To create \vcells, the mobile device collects topology fingerprints and progressively aggregates scans based on a commonality metric.

\textbf{General Design.} To collect our datasets we used Hunter\footnote{Hunter App is an Android application developed at the University of Los Andes by Andr\'es Arcia-Moret and Jose Marquez, available at \texttt{goo.gl/6QHUiR}}, an Android mobile phone application that scans WLAN topologies. We used this application to store Wi-Fi AP data and use those data points to map APs into \vcells to predict places of mobility and interaction.

 The core of the application is composed of helping modules that support the \vcell calculation, namely the GPS location (for providing the ground truth), Wi-Fi scanning module and path approximation algorithm (to adjust the geo-position based on existent maps). From a user's perspective, a mobile can collect the surrounding WLAN topology to interact conveniently with a central service (e.g., a Topology Manager, see \cite{arciamoret2,arciamoret3} for further information). 
 
By default, the scanning module is meant for two main purposes: topology discovery and keeping alive a session between a mobile and its serving AP. In our particular case, the scanning function can be conveniently and continuously invoked (and with different invocation frequencies) to obtain different perspectives of the surrounding topology. Since the full-scaning time varies on every mobile device, we suppose a central service that helps to build \vcells (we discuss a similar service in \cite{arciamoret2,arciamoret3}).

\textbf{\vcell algorithm.} We cluster or group APs by AP detection frequency. The \vcell formation algorithm is defined in Algorithm~\ref{list:cells}. The cell size, i.e., how many scans form a cell, is controlled by the \textit{cell condition} (CC). The CC is an adjustable commonality metric that defines the quantity of overlapping APs within a \vcell. So, if the intersection between the current scan and the tentative \vcell (\texttt{\small curr\_cell}) satisfies the CC, then the \texttt{\small scan} will be part of the \texttt{\small curr\_cell}. Essentially, the \texttt{\small unique()} function ignores duplicates between \texttt{\small curr\_cell} and \texttt{\small scan}. If the CC is not satisfied then the \texttt{\small curr\_cell} is appended to the \texttt{\small vcellList}, and a new \vcell will get formed during next iterations. The algorithm terminates when all \vcells are defined. Note that one AP can be a part of many cells, thus forming intersection areas.

Algorithm~\ref{list:overlap} shows how we define intersection areas. The \textit{overlap condition} is an adjustable commonality metric that is a range representing overlap among \vcells. During the \vcell formation process, the overlapped geo-coordinates should be close enough to have sufficient APs. The quantity of APs is controlled by the \textit{overlap condition} (OC). For each \vcell we check all other \vcells and if the OC bounds the unique intersection between any two \vcells, then we append the two together in the \texttt{\small overlapList}. This can be a very costly operation, so in practice, we restrict the overlap of each \vcell to its adjacent \vcells.


During the \vcell construction process, we use geo-coordinates for defining an overlap area. As we have mentioned in Section~\ref{sec:introduction}, such a technique can be used during an initial topology recognition process to bootstrap the \vcell database. Later on, this database can be regularly updated with incremental contributions of users (as in \cite{arciamoret3}). However, the overlap (Algorithm~\ref{list:overlap}) is still optional and does not affect how \vcells are formed (Algorithm~\ref{list:cells}). Initially, we plan to use a couple of examples to evaluate \vcells created using the algorithm: running speed and walking speed. See Section~\ref{sec:evaluation} for algorithm evaluation and how \vcell properties vary in different scenarios.

\begin{algorithm}
\footnotesize
\caption{\vcell formation}
\begin{algorithmic}
\REQUIRE $APs\ scan\ logs (scanList)$
\STATE $vcellList \leftarrow vcell\ from\ first\ scan\ of\ scansList$
\STATE $curr\_vcell \leftarrow vcellList.pop()$
\FOR{$each\ scan\ in\ scansList$}
\IF{$(scan \cap curr\_vcell) \geq CC*len(curr\_vcell)$}
\INDSTATE[4] $curr\_vcell \leftarrow unique(curr\_vcell \cup scan)$
\ELSE
\INDSTATE[4] $vcellList.append(curr\_vcell)$
\INDSTATE[4] $curr\_vcell \leftarrow scan$
\ENDIF
\ENDFOR
\end{algorithmic}
\label{list:cells}
\end{algorithm}

\begin{algorithm}
\footnotesize
\caption{Define overlap between vcells}
\begin{algorithmic}
\REQUIRE $vcellList$
\FOR{$i\ in\ vcellList$}
\FOR{$j\ in\ vcellList$}
\IF{$(vcellList[i] \cap vcellList[j])\ is\ bounded\ by\ OC$}
\STATE $overlapList \leftarrow (vcellList[i] \cap vcellList[j])$
\ENDIF
\ENDFOR
\STATE $vcellList[i] .append(overlapList)$
\ENDFOR
\end{algorithmic}
\label{list:overlap}
\end{algorithm}

\textbf{Bloom filter assisted device location discovery.} Two requirements are justifying this component: (1) simplicity in the implementation of a location discovery service, and (2) a CPU and energy efficient search of a fingerprint in the \vcell database. A BF is an ideal data structure to cover these requirements because a BF can be quickly built out of a small number of scans, then matched up with an established \vcell. However, false positives (FP) are an inherent characteristic of BFs. In our particular case, during a journey when \vcells have already been formed, an FP can be detected if the fingerprint starkly differs from previous \vcell locations. These include abrupt jumps in location or insensible fingerprint information. Hence, we can proceed to correct such instances. According to the probability equation of false positives \begin{math} p = (1 - [1 - 1/m]^{kn})^k \end{math}, a BF with 1.1 million entries and 7 hash functions has about a 0.5\% chance of a false positive. The chances are that for very populated \vcells, bounded by a maximum of few hundreds of APs and with as many hash functions, FPs are almost negligible.




\section{Evaluation}
\label{sec:evaluation}


We have used the Hunter application to perform two representative journeys varying the speed of data collection. We have performed one collection in the downtown of M\'erida city in Venezuela, at walking speed (5 kph) in which we have found 1826 APs for a 5 km journey. And a similar one in the Alexandras Avenue in Athens, Greece at running speed (10 kph) in which we have found 687 APs for 15 km journey. Both measurements were performed during June 2015 and using high full-scaning times. Collected lists are conveniently passed to the central service by doing delay-tolerant uploads as in~\cite{shi}. As expected, the distance between two different consecutive scanning is more spaced when the speed of collection is higher. 

\begin{figure}
	\subfigure{{\includegraphics[width=0.49\linewidth]{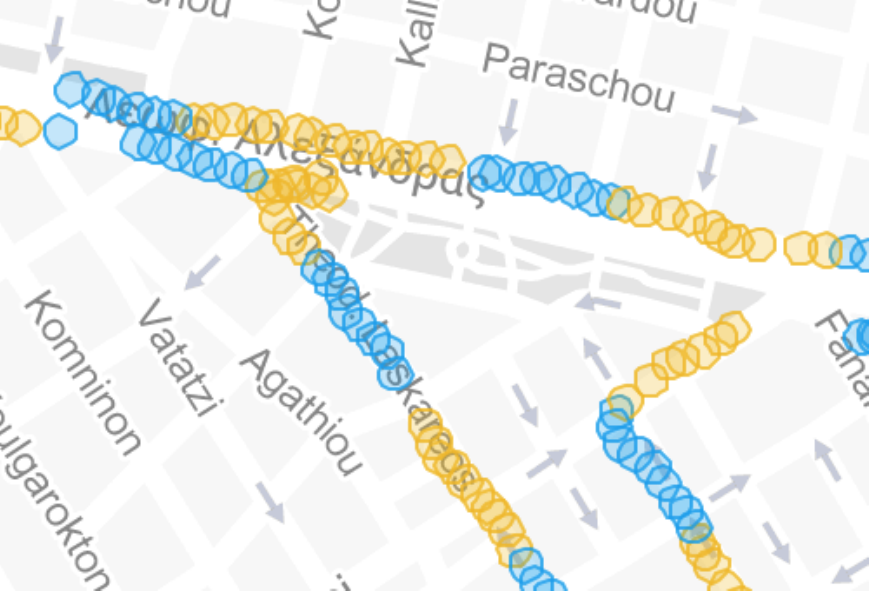}}~ {\includegraphics[width=0.49\linewidth]{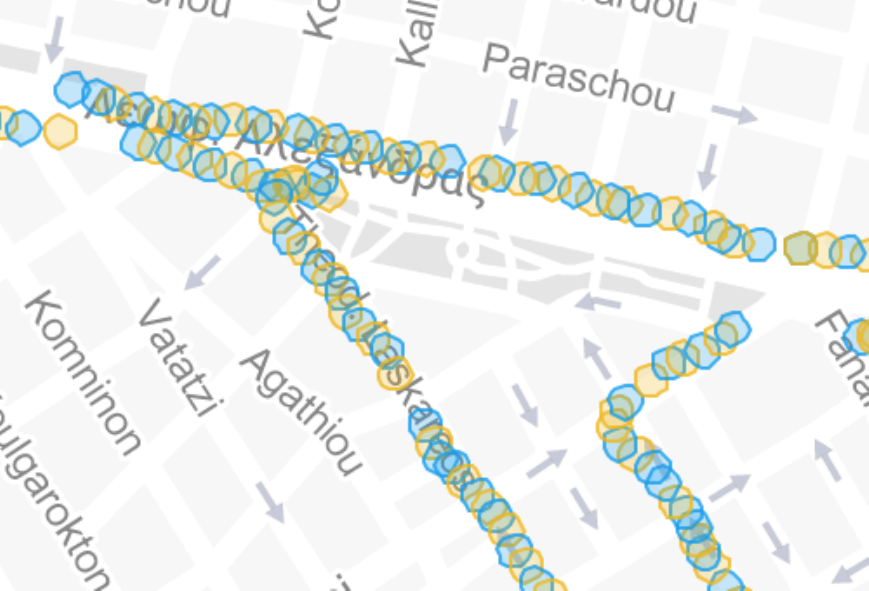}}}
	\vspace{-0.4cm}
	\caption{Athens with cell condition 0.20 (left) and 0.80 (right)}
	\label{fig:athens}
\end{figure}

Fig.~\ref{fig:athens} presents different cell conditions produced by Algorithm~\ref{list:cells} for the journey in Athens. One possible application for different \vcell sizes corresponds to the refined location. As the cell condition gets more relaxed, for example as depicted in Fig.~\ref{fig:athens} (left), cells become bigger, and it is easier to find certain fingerprint inside the cell. This rough approximation could be refined through other less relaxed conditions, such as depicted in Fig.~\ref{fig:athens} (right). Varying the precision of the cells could have multiple uses, the location of a mobile could be more precise when using smaller cells. Actually, from the collected datasets one can see that small cells (for both speeds of collection) approximate the location of a fingerprint to a few dozens of meters, much like a GPS position.

On top of the previous application, we could infer the mobility pattern of an individual device. Depending on the needs we could approximately indicate whether a mobile is going through a particular sequence of cells, or (using more relaxed conditions) the mobile is indeed positioned within a large \vcell. This information is useful in use-cases such as estimating the dynamics of disease propagation, being especially helpful in developing regions.

\begin{figure}
    \centering
    {\includegraphics[width=0.38\textwidth]{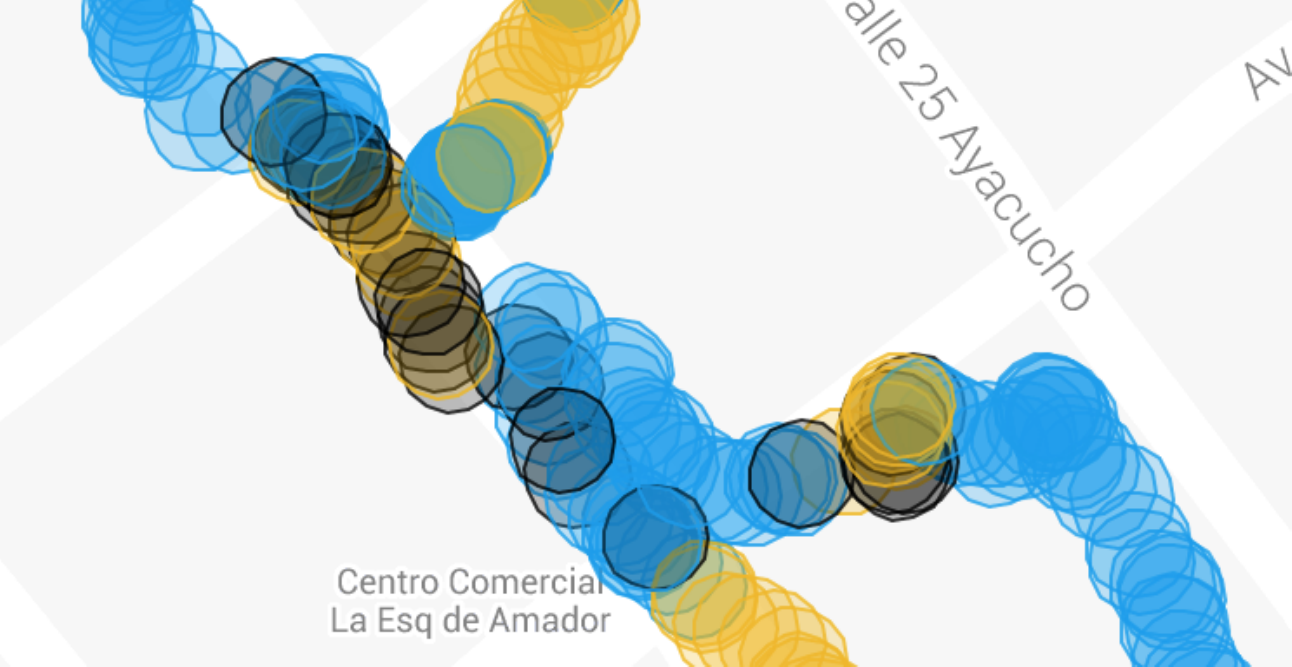}}
    \caption{Cell condition 0.30 and overlap condition [0.20, 0.30] for a fast-scanned and densely covered urban area.}
    \label{fig:merida302029}
\end{figure}

Finally, Fig.~\ref{fig:merida302029} illustrates the case for a cell overlapping. Cell overlaps is an indication of the continued coverage provided by the spontaneous deployment within the urban area. We have tested both relaxed and more strict cell conditions, and results have been similar. There are always overlapping scans whenever a cell condition is relaxed. For the case shown in Fig.~\ref{fig:merida302029}, we have set the cell condition to 0.30. The black markers show a common subset of scans between two neighbor cells indicating a transition when walking from one cell to the next. The common subset of scans reports an overlap condition between $20\%$ and less than $30\%$ of the quantity of APs shared among the two cells.

 \hfill

There are a few inherent shortcomings of the system that should come to light. The scanning process during data collection is challenged by various obstacles, depending their shape and size. Typically obstacles include people, vehicles, and buildings. All these elements compose particular environments that alter scanning results. We have also observed that, in more open spaces, i.e., having fewer obstacles, \vcell sizes tend to be bigger for a fixed cell condition. Moreover, at a higher speed of collection (but at the same scanning frequency), \vcells tend to be less dense, i.e., it results in a fewer APs per fingerprint. However, for our test cases, Algorithm~\ref{list:cells} always builds a continuum of \vcells.

\section{Related Work}
\label{sec:related}

Mobility has been widely studied using geo-localised cellular connectivity data for a variety of different reasons~\cite{ Trestian2009, Finamore2013, Noulas2015}. While cellular data covers macro regions and is extremely rich in spatiotemporal information, Wi-Fi data is confined to micro-regions and can provide better insights into problem spaces that cannot be addressed using cellular data.

In \cite{piotr} experiments recording up to 6 months of human mobility data have been conducted, with high temporal resolution, finding a strong correlation between simple Wi-Fi scanning traces and human location. Their results encourage the use of Wi-Fi positioning through simple scanning wardriving due to the inferred high presence (on crowded Wi-Fi places) of humans in their sample. Although there exist other studies about Wi-Fi wardriving and location prediction systems \cite{rekimoto, kawaguchi}, they do not address any mechanism to convert scans systematically into meaningful positions.

Location prediction through Wi-Fi network has been studied from many perspectives, but to the best of our knowledge, none of them address the problem of clustering APs into \vcells as we discuss in this article. \cite{muthukrishnan} uses Wi-Fi AP location and characteristics along with GSM to predict the speed of the mobile device. They have studied the duration of the strongest Wi-Fi access point perceived. In their particular case, authors demonstrate that different degrees of mobility (dwelling, walking or driving) can be inferred through the collection and post-processing of Wi-Fi and GSM beacons. \cite{song} uses Layer-3 location prediction by processing Wi-Fi traces. They present an analysis based on Markov models on users location from extensive access point interactions. They define a single location of a user as an established connection (i.e., scanning, authentication, and association, and later disconnection) to a single AP during a certain period. They rely on user attachment and detachment traces on the confined network from which authors have complete access to the extensive logs.

Traceability through Wi-Fi technology has been investigated from several end-point perspectives. In \cite{rose}, the authors propose a cooperative system called Argos capable of distributively collect Wi-Fi traffic in an urban area. A group of geographically distributed sniffers is instructed to be aware of a desired client unique address, and after the required traffic is collected, all intervening sniffers pass the information to a central server which in turn infers about the mobility patterns. 

Finally, there has been a considerable amount of literature claiming the use of the received signal strength (RSS) as the primary indication of the position of a mobile \cite{miao, kaemarungsi, saxena}. However, these approaches refer to a physical layer procedure based on propagation models to infer the approximative location of a particular device. There is a broad range of variations of this location prediction mechanism and it is tough to follow a single one that fits all devices. Moreover, the RSS depends entirely on the receiving antenna and radio so that the result can vary within two different mobiles in the same location. Furthermore, to improve the estimation of the position, generally large amount of data is required, thus increasing the energy consumption \cite{sugano}.

\section{Conclusion}
\label{sec:conclusion}

We have introduced a novel technique to organise APs found in spontaneous deployments into what we define as \vcells. Different from existing Wi-Fi location techniques, \vcells are built from progressive Layer 2 active scanning. We have explained \vcells and their properties such as their shapes, associated properties and proposed a way of defining and exploiting overlapping of \vcells. For this, we have collected data from deployments in urban areas, and we have illustrated \vcells emulating as typical vermiform shapes considering different mobile reception conditions.

We have discussed a novel geolocation system for mobile devices using Bloom filters to represent \vcells and to provide a simple position retrieval method. We have argued that BF representation could be potentially be exploited for scenarios in which the abstraction of mobility patterns is useful, thus making \vcells relevant for localisation amidst disasters, human tagged locations, and mitigating the lack of cellular coverage.

\end{document}